for Paper I.

\documentstyle [12pt] {article}
\centerline{\large\bf SIMULATION OF THICK ACCRETION DISKS }
\centerline{\large\bf WITH STANDING SHOCKS}
\centerline{\large\bf BY SMOOTHED PARTICLE HYDRODYNAMICS}

\vspace{0.05cm}

\centerline{Diego Molteni}
\centerline{\small University of Palermo, 90100 Palermo, Italy}
\vspace{0.01cm}
\centerline{Giuseppe Lanzafame}
\centerline{\small Osservatorio Astronomico, University of Catania, Catania,
Italy}
\vspace{0.01cm}
\centerline{Sandip K. Chakrabarti$^*$}
\centerline{\small Astronomy and Astrophysics}
\centerline{\small The University of Chicago, 5640 S. Ellis Av.,
Chicago, Il, 60637}
\vspace{0.00cm}
\centerline{and}
\vspace{0.00cm}
\centerline{\small Aspen Center for Physics, Aspen, Colorado 81611}

\begin{document}
\pagenumbering{arabic}

\newpage

\centerline{\bf Abstract}

\baselineskip 24pt

{\small We present results of numerical simulation of inviscid
thick accretion disks and wind flows around black holes. We use
Smoothed Particle Hydrodynamics (SPH) technique for this purpose.
Formation of thick disks are found to be preceded by shock waves
travelling away from the centrifugal barrier. For a large range of
the parameter space, the travelling shock settles at a distance
close to the location obtained by a one-and-a-half dimensional model of
inviscid accretion disks. Occasionally, it is observed that accretion
processes are aided by the formation of oblique shock waves, particularly
in the initial transient phase. The post-shock region (where infall velocity
suddenly becomes very small) resembles that of the usual model of thick
accretion disk  discussed in the literature, though they have considerable
turbulence. The flow subsequently becomes supersonic before falling into the
black hole.
In a large number of cases which we simulate, we find the formation of strong
winds which are hot and subsonic when originated from the disk surface
very close to the black hole but become supersonic within a few tens of the
Schwarzschild radius of the blackhole.
In the case of accretion of high angular momentum flow, very little amount of
matter is accreted directly onto the black hole. Most of the matter is,
however, first squeezed to a small volume close to the black hole, and
subsequently expands and is expelled as a strong wind. It is
quite possible that this expulsion of matter and the formation of cosmic radio
jets is aided by the shock heating in the inner parts of the accretion disks.}

{\it Subject Headings}:\ {\rm accretion, accretion disks - black hole physics
-  hydrodynamics - shock waves \\}

* On temporary leave of absence from Tata Institute of Fundamental Research,
Homi Bhabha Road, Bombay-400005, INDIA (Permanent Address)\\

\newpage

\noindent {\large 1. INTRODUCTION}

\baselineskip 24pt

In the literature, some results of the numerical simulation of thick accretion
disks using finite difference methods are present
(Wilson, 1978; Hawley, Smarr \& Wilson, 1984, l985). These works show that the
thick accretion disks could form during the accretion of matter which has a
significant angular momentum. It is found that shocks are formed which travel
outward in the disks. This work also points out
that 'hollow' jet like features are produced which propagate roughly
along the funnel wall. Subsequently, Eggum, Coroniti
and Katz (hereafter ECK, l987, 1988) simulated  Keplerian disks with viscosity
and radiation transport. These solutions do not produce shock waves,
but strong winds are produced. In paper I, (Chakrabarti \& Molteni, l993) we
presented basic motivations of studying disk models which include shock
waves as well as some results of one-dimensional simulations
showing that in an inviscid thin disk {\it standing shocks} should be quite
common.

In the present paper, we simulate the formation of thick disks, keeping
in mind that not only can travelling shocks form as noted
by earlier workers, but {\it standing shocks} can form as well. In view of
the recent advances in understanding of shock formation in the `thick' disk
(see, Chakrabarti 1989, 1990a), we have {\it a priori} knowledge of the
parameter space (spanned by, say, the energy ${\cal E}$ and angular momentum
$\lambda$ of the flow) for which shocks may or may not form in the disk.
Guided by the input from this analytical work (which is, at best,
valid for one-and-a-half dimensional inviscid disks, where the vertical
motion is neglected compared to the other velocity components), we have been
able to understand in a fairly complete manner the status of shock formation
in thick accretion disks as well as the manner in which winds may
originate. Our principal conclusions are: (a) for a
significant range of energy and angular momentum of the flow, the
thick disks may contain standing shock waves; (b) models which
ignore the vertical motion inside the disk are reasonably good,
particularly in the preshock region of the flow; (c) shock locations predicted
by the one-and-a-half dimensional models are at lower radial distances than
the shocks locations actually obtained in the simulations -- this discrepancy
is removed when the effects of the turbulent pressure in the post-shock region
and the extra compression due to the vertical motion are included;
(d) of the two shock locations predicted by the analytical work, the
shock located farther away from the black hole is stable -- this conclusion is
similar to that obtained for one dimensional thin accretion flows
[paper I];(e) there could be considerable mixing in the matter
in the postshock region near a
black hole; and finally, (f) a strong supersonic wind can originate from
regions close to the black hole, and the shock heating in this region is
probably an essential ingredient for such behavior.

A major difference in the manner in which shocks form in a thin
disk (Paper I) and a thick disk (of the present work) is that whereas
in the thin (one dimensional) disk, some perturbation must
be introduced in order to have a shock, in the thick disk, perturbations
are always present in the form of turbulence. Thus, we do not obtain
any solution which does not contain shocks, stationary or non-stationary.
Our work also brings about a major departure from the standard models of
disks in astrophysics. From the results of paper I and the present paper,
we claim that shock waves are more common in accretion disks than hitherto
realised, and their effects must be included in interpreting observed
results from AGNs and stellar disks. This statement is particularly true
when the disk is radiation pressure driven rather than viscous driven,
i.e., when the infall timescale due to pressure gradient force is shorter
compared to the timescale of the transport of angular momentum by viscosity.

The plan of the present paper is the following: In the next Section, we
briefly present analytical models of thick accretion disks which
contain significant radial motions (unlike the canonical thick disk
models of, say, Paczy\'nski and Wiita, 1980, where motions other than
that in the azimuthal direction are ignored). These models
can be found in detail in Chakrabarti (1989, 1990a). However
we now qualitatively include the effects of the turbulent pressure in
the post-shock region and the extra compression due to significant vertical
velocity component.  In \S 3, we produce results of a few numerical
simulations. Finally, in \S 4, we summarize our results and make concluding
remarks.

\noindent{\large 2. SHOCK LOCATIONS IN A 1.5D FLOW WITH TURBULENCE}

We assume a rotating, axisymmetric, adiabatic, accretion flow in
vertical equilibrium, near a black hole.
We take Newtonian models for the non-rotating central compact object
as given in terms of the Paczy\'{n}ski \& Wiita (1980)
potential. We also assume a polytropic equation of state for the
accreting (or, outflowing) matter, $P=K \rho^{\gamma}$, where,
$P$ and $\rho$ are the isotropic pressure and the matter density
respectively, $\gamma$ is the adiabatic index (assumed in this
paper to be constant throughout the flow, and is related to the
polytropic index $n$ by $\gamma = 1 + 1/n$) and $K$ is related
to the specific entropy of the flow $s$: $s = const$ implies $K = const$.
We assume that the flow is non-dissipative, so that the specific
angular momentum $\lambda$ is constant everywhere. This also implies that
the entropy density, and thus $K$ can vary  only at the shock.
A complete solution of the stationary model requires the equations of energy,
angular momentum and mass conservation supplied by transonic conditions at
the critical points and the Rankine-Hugoniot conditions at the
shock. The general procedure followed is the same as is presented
in Chakrabarti (1989, 1990a). Presently, however, we include the
qualitative effects of the turbulent pressure in the post-shock flow in order
to obtain shock locations in the thick, turbulent disk. When the vertical
component of velocity is significant, its infall toward the equatorial plane
causes an extra compression of the flow. Effects of this has also been
qualitatively included. Inclusion of these effects show that the parameter
space in which shocks can form may be vastly enhanced than what is described
in Chakrabarti (1989, 1990a).

In what follows, we use the mass of the black hole $M$, the velocity of
light $c$ and the Schwarzschild radius $R_g=2GM/c^2$ as the units of
mass, velocity and distance respectively. The dimensionless energy
conservation law can be written as,
$$
{\cal E} = \frac{\vartheta^2}{2}+\frac{a^2}{\gamma-1}
+ \frac{\lambda ^2}{2x^2}-\frac{1}{2(x-1)}
\eqno{(1a)}
$$

Here, $\vartheta$ and $a$ are the non-dimensional radial velocity and  sound
speed, $x$ is the non-dimensional radial distance. Apart from an unimportant
geometric factor, the mass conservation equation is given by,
$$
{\dot M} = \vartheta \rho x h
\eqno{(1b)}
$$
where $h$ is the constant half-thickness of the flow which, assuming
hydrostatic equilibrium in vertical direction, is
given by $h=a x^{1/2} (x-1)$.
It is useful to write the mass conservation equation in terms of
$\vartheta$ and $a$ in the following way,
$$
\dot{\cal M} = \vartheta a^{2n+1} x^{1/2} (x-1)
\eqno{(2)}
$$
We shall use the phrase
`accretion rate' for this quantity, keeping in mind, however, that
${\dot {\cal M}}\sim {\dot M}{K^n}$ does not
remain constant at the shock because of the generation of entropy.
The shock conditions which
we employ here are the following (subscripts `-' and `+' refer to
quantities before and after the shock): The energy conservation equation,
$$
{\cal E}_+ = {\cal E_-},
\eqno{(3a)}
$$
the pressure balance condition,
$$
W_+ +  W_{T+}+ \Sigma_+ \vartheta_+^2 = W_- + \Sigma_- \vartheta_-^2
\eqno{(3b)}
$$
and the baryon number conservation equation,
$$
\dot M_+ = \dot M_-
\eqno{(3c)}
$$
Here, $\Sigma$ and $W$ are the density and pressure respectively
which are integrated in the vertical direction (Chakrabarti, 1989).
In Equation 3b, we include turbulent pressure $W_T \sim \alpha \Sigma  a^2$,
$\alpha$ is assumed to be a constant ($ < 1$) measuring the effects
of turbulence. A part of $W_T$ is contributed by
an extra compression of the flow due to the infalling matter from
the vertical direction, as the vertical velocity could be significant.
In order to have a shock, the radial flow must be supersonic, i.e., the
stationary flow must pass through a sonic point. The sonic point
conditions are derived in Chakrabarti (1989) and will not be discussed here
in detail. However, we derive here the Mach number relation, which
enables one to obtain the shocks locations very easily.

{}From Equations 3(a-c) and (2), at the shock location $x=x_s$, we obtain,
$$
\frac{1}{2}M_+^2 a_+^2 + \frac{a_+^2}{\gamma-1}=
\frac{1}{2}M_-^2 a_-^2 + \frac{a_-^2}{\gamma-1}
\eqno{(4a)}
$$
$$
\dot{\cal M}_+ = M_+ a_+^{2n+2} x_s^{1/2} (x_s-1)
\eqno{(4b)}
$$
$$
\dot{\cal M}_- = M_- a_-^{2n+2} x_s^{1/2} (x_s-1)
\eqno{(4c)}
$$
$$
\frac{a_+^{2n+3}}{\dot{\cal M}_+} [\frac{2}{3\gamma -1} + (\alpha+M_+^2)]
=
\frac{a_-^{2n+3}}{\dot{\cal M}_-} [\frac{2}{3\gamma -1} + (\alpha+M_-^2)]
\eqno{(4d)}
$$

After elimination of some variables, we obtain the
relationship between the pre-shock and the post-shock mach numbers,
as given by,
$$
M_+M_-=\frac{2+\alpha(3\gamma-1)}{(3\gamma -1)^2 - C (\gamma - 1)}
\eqno{(5)}
$$
in terms of the Mach invariant function, $C=C(M)$ at the shock,
$$
C=\frac{[2+(3\gamma -1)(\alpha + M^2)]^2}{M^2 [2+(\gamma -1)M^2]} .
\eqno{(6)}
$$

Figure 1 shows the parameter space spanned by the specific energy
${\cal E}$ and specific angular momentum ${\lambda}$ of the flow
for which standing shocks may be formed in the disk when
$\alpha=0$ is chosen (Chakrabarti 1989, 1990a). This region,
bounded by three curves, is lebelled with $\Sigma$. Numerically, however,
standing shocks were found even outside this region, because
in general, some turbulence is always present in the thick disk. In Figure 2,
we show the effects of turbulence on the location of the shocks
($X_{s3}$ of Chakrabarti 1989) for $\lambda=1.65$
for a range of $\alpha$ and specific energy  ${\cal E}$.
It is clear that shocks form farther from the hole, as the turbulent
pressure as well as the extra compression due to vertical motion
goes up. Another important point to note is that the range
of energy for which shocks form with $\alpha\ne 0.0$ is much
higher than the range obtained with $\alpha=0.0$. This shows that
the chance of shock formation is much higher in a turbulent disk, particularly,
for low enough energy.

\noindent {\large 3. RESULTS OF SIMULATION OF THICK ACCRETION DISKS AND WINDS}

The Smoothed Particle Hydrodynamics (SPH) method that we use has been primarily
developed to deal with fluid dynamics in astrophysical context
(Lucy 1977; Gingold \& Monaghan 1977; for a recent review, see, Monaghan,
1992). In Paper I, we have presented the procedure for the
implementation of the code in axisymmetric cylindrical co-ordinates.
The results described in this Section are obtained with a range of
initial conditions which may prevail in  realistic circumstances.
We study the following diverse cases: (A) the injection rate is uniform with
height at the outer edge and (B) the injection rate is such that at the
outer edge, the disk vertical structure is isothermal. We simulate (B)
with low as well as high angular momentum. Since in this case, the
flow is in equilibrium at the injection radius, the infall is less
`violent' than in case (A), and it takes longer
time to reach a steady state solution. Due to limitations in computing
time, we had to stop the simulation in case (B) before a {\it complete}
steady state is reached, though the simulations were carried out to
times much longer than the infall timescale. In all the cases, we
inject matter only in one quadrant and use reflection boundary
condition to obtain the solutions in other quadrants.

We note here that ${\dot{\cal M}}$ (but not ${\dot M}$) is an eigenvalue of
our problem and is fixed by a choice of the input parameters
${\cal E}, \lambda$. We therefore chose the density of matter at the outer
edge to be unity and the rate of matter injection is automatically
adjusted to achieve a steady state flow. The reference density
is obtained by the actual accretion rate ${\dot M}$. For an accretion
rate of ${\dot M}={\dot m} {\dot M_{Edd}}$, the reference density is
$$
\rho_{ref} = \frac{{\dot m}{\dot M_{Edd}}}{4 \pi c X_{ref}^2}
\eqno{(7a)}
$$
where, $X_{ref}=\frac{2GM}{c^2}$, and the reference temperature is,
$$
T_{ref}=[3 c^2 \frac {\gamma -1}{a}\rho_{ref}]^{1/4}
\eqno{(7b)}
$$
with $a$ as the radiation constant.

Figures 3(a-d) show results of a Case (A) simulation.
Particles with angular momentum $\lambda=1.65$ and energy ${\cal E}=0.006$
are injected at the outer edge of the disk at $x=30.0$. The result
shown is at $T=700$ when the flow has achieved steady state. The total
number of particles is $60,000$. A shock
is clearly formed at $x\sim 16.6$. Using the 1.5D model without any
turbulent pressure ($\alpha=0.0$) the shock location for these
parameters is at $X_{s3}=11.2$,
which is much closer to the black hole than the location
we observe here. The post-shock flow clearly contains
by a strong vertical motion (ignored in the 1.5D model) as well as
turbulence. Assuming the shifted location of the shock is entirely
due to the turbulent pressure, we obtain $\alpha=0.14$, which is
very realistic (see, Fig. 2). A weaker oblique shock is also produced
in this example. Later we will show a stronger case of oblique
shock formation (see, Fig. 5b below).

In Figure 3a, X-Z locations of the `pseudo'-particles are presented.
In Figure 3b, we zoom a region closer to the hole, in order to show
detailed behavior of the flow. We plot arrows at one in every five
particles for clarity. The length of an arrow
is proportional to the Mach number of the flow at the location where
the arrow is originated ($M=1$ corresponds to $0.2$ in length). A
few observations could be made in this context: (a) The shock is very sharp
(width of about $0.2-0.4$ Schwarzschild radius), (b) The subsonic flow
in the post-shock region becomes highly supersonic before plunging
into the black hole, (c) The postshock  flow is first
diverted away from the equatorial plane
due to the presence of a very strong turbulence between $r=7.0$ and $r=14.0$
and then enter into the hole at an angle.
(d) A part of the flow is diverted altogether from the black hole in the
form of winds. This latter behavior is due to the heating
and subsequent expansion in the postshock region. The overall
behavior of the Mach number is clear in Figure 3c, where the contours of
constant Mach number are plotted. The contours are drawn
in the linear scale with minimum, maximum and interval  given
by: $M_{min}=0.03$, $M_{max}=2.0$ and $\Delta M=0.2$. Note in particular
that the wind becomes supersonic at distances as close as $x\sim 22$.
In Figure 3d, we show the contours (in linear scale)
of contant temperature (dimensionless) of the flow with
$T_{min}=0.14$, $T_{max}=1.2$ and $\Delta T = 0.06$. Note that
the contours of constant temperature resemble those of canonical
thick disk models (cf. Paczy\'nski and Wiita, 1980) in the post shock
region. This is due to
the fact that in the immediate vicinity of the postshock region,
the radial velocity is very small, and our simulation reproduces the original
thick disk solution. Our result indicates that the postshock
region of the flow could be considered as the thick accretion disk.

In the above example, we showed only the final result. At initial stage
of the simulation the shock is formed very close to the hole,
which subsequently travelled outward to reach its stable location.
This behavior is very similar to what was observed in Paper I.
Similarly, by reducing the vertical height of the disk, we recover
the result that the shock at $x\sim 16.6$ actually corresponds to $X_{s3}$,
the outer shock of the analytical solution of Chakrabarti (1989). Thus,
as in Paper I, only the shock at $X_{s3}$ is found to be stable.

In Case (B) simulations, matter is assumed to be in hydrostatic
equilibrium in the vertical direction at the outer edge. Figure 4(a-d)
shows the result of a simulation for $\lambda=1.625$, ${\cal E}=0.007$
where isothermal vertical structure at the outer edge is used.
This case is so chosen that the parameters lie at the boundary between shock
and no-shock solutions in analytical 1.5D model (with $\alpha=0$).
At a little higher energy a weak shock should form at around $X_{s3}=6.9$.
The observed shock in the simulation at $\sim 14.0$ corresponds
to $\alpha=0.2$ in our present model. The parameter space for which
shocks may form is vastly bigger in presence of turbulence and
vertical motion in the disk. This behavior is found to be particularly
significant when low angular momentum flow is considered. This is because
the turbulent pressure becomes comparable or more than the
centrifugal pressure of matter close to the black hole.

In the next example, we study the formation of winds, as well as the
interaction of winds with the accreting matter.
We choose the injection radius at $x=30$ and the wind formed is allowed to
go much beyond this radius. The thickness of the disk at the outer
edge ($x=32$) is chosen to be such that the density falls off to a value
of about $10$ percent of the equatorial density.
Figure 4a shows the flow behavior. Length of the
arrows are proportional to the local Mach numbers. Note that a shock is formed
at about $x=14$ where the flow becomes subsonic, and another shock
is formed in the region where the wind is interacting with the injected
matter. The interaction heats up the gas and matter is turned around
to join the strong wind that is developed above the disk.
Figure 4b shows contours of constant Mach number in linear scale.
The minimum, maximum and the interval of Mach number are: $M_{min}=0.04$,
$M_{max}=4.0$, $\Delta M = 0.2$. Notice that the shock strength
(namely, Mach number jump at the shock) at the interaction region is much
higher compared to the shock on the equatorial plane. Also note that the wind
formed certainly becomes supersonic by $x\sim 30$. In Figure 4c, we show the
contours of constant temperature in linear scale with $T_{min}=0.076$,
$T_{max}=0.8$ and $\Delta T=0.02$. As in Figure 3d, the temperature
distribution in the post shock flow also resemble
as that in the thick accretion disks of Paczy\'nski and Wiita (1980).
In Figure 4d, we show the ratio of the outflow rate to the injection rate
as a function of time. The outflow commences at $T\sim 400$ and steadily
rises to about $10$ percent.
The rapid fluctuation throughout is mostly due to noise, typical of SPH
technique. We stop the simulation at about $T=2000$, at which time
the outflow ratio was slowly increasing.

We present another Case (B) simulation where we use a high angular
momentum ($\lambda=1.80$). The objective was to simulate disks which
are much larger as well as thicker. We inject matter at $x=100$ and
the thickness chosen to be $x=70$ at this edge. The disk is vertically
cut off at $x=70$ where the density is $10$ percent of the equatorial density.
Because of limitations on computer processing time, we stop the simulations
at $T=1780$ when there are $63,345$ particles in the region of
integration. The number was still increasing very slowly, indicating
that the steady state was not reached yet.
Because of high angular momentum, shocks are produced at a very large
radius ($X_{s3}=50.0$ in the 1.5D model with $\alpha=0$).
There is a considerable backflow of the matter (bounced off the centrifugal
barrier close to the hole) which diverted the flow off the equatorial plane.
Interaction with the diverted flow with the infalling matter
from a very large vertical height causes oblique shocks to develop. Figure 5a
provides the Mach number distribution in this case, showing
a vertical strong shock forming at about $x=40$. Oblique shocks are
also seen. In Figure 5b, we zoom the region with triple shocks which
are remarkably resolved in our simulation. The flow coming
closer to the equatorial plane is diverted away by the first shock,
but a second oblique shock is seen to refract matter toward the black hole.
A weak backflow near the
equatorial plane is also seen. Such backflows were also present in
the initial transient phase of the Case (A) simulation (Figure 3).
However, it developed into small scale turbulence when otherwise steady
state was reached. We expect that similar behavior will prevail in the present
case also.

\noindent {\large  5. SUMMARY AND CONCLUDING REMARKS}

In this paper, we have studied the nature of the Rankine-Hugoniot
shocks in thick accretion disks using Smoothed Particle Hydrodynamics
method. Our numerical simulations indicate that shocks are essential
ingredients in the formation of a thick accretion disk, and that
they could be present for a large range of initial parameters.
Unlike simulations by earlier workers, we find disks with
stationary shocks. The initial injection condition at the outer edge
was chosen from the 1.5D analytical models
and location of the shock appears to be roughly in agreement with the
prediction from 1.5D model, though, inclusion of the effects of
turbulence in the post shock region as well as the extra compression
(both of which we model qualitatively here) renders a better agreement between
the numerical and the theoretical results. In Paper I, we noted that the
outer shock at $X_{s3}$ (Chakrabarti, 1989) was chosen by the flow
and not the one located at $X_{s2}$. We find similar result in our
present 2D simulation also. An important difference, however, is that,
whereas, in the one dimensional simulations (Paper I), one had to
use `perturbed' initial condition in order to produce shocks, in the
two dimensional simulations, the turbulence present is sufficient
to `induce' shock formation. Thus shocks are always present in our
simulations even when the parameter space is so chosen that the
they are non-stationary.

Our observations regarding shock formation inside the disk
are particularly true when the viscosity of the disk is very low
and the flow is mostly pressure driven rather than viscous driven.
Chakrabarti (1990b), in his study of
standing shocks in viscous isothermal disks, found that the shock
located at $X_{s3}$ becomes weaker (and ultimately disappear)
as viscosity is increased.
This result, together with the present simulation, indicates why
in the standard models of the accretion disks where
the flow is viscous driven (Shakura \& Sunyaev 1973) one need not be
concerned with shock formation. Indeed, the simulation
of viscous, {\it initially Keplerian, axisymmetric accretion disks}
(as opposed to our constant angular momentum disk) by ECK (1987, 1988) does
not show any shock waves. In this case, viscosity reduces the angular momentum
significantly to sub-Keplerian values throughout the disk
[see, Fig. 10 of ECK (1988)] so that the flow faces a very weak
centrifugal barrier. The flow does not strongly `bounce back' at the barrier
and as a result, the winds are found to have much weaker
kinematic flux. Winds in these simulations carry a fraction of a percent
of the infalling matter (as opposed to our simulation where a sometimes
almost ten percent of matter is blown away) and the
specific angular momentum carried by the wind is smaller as well.
These simulations show strong, non-steady  equatorial
{\it outflows}, the physical origin of which are not immediately obvious.

A major problem in accretion disk theory is to provide a suitable
prescription for viscosity mechanisms which may be
operating in the disk. Recently, a significant progress has been made
in the literature. It appears that non-axisymmetric spiral waves (Spruit,
1987; Hanawa, 1988; Chakrabarti, 1990c) or internal waves (Vishniac \&
Diamond, 1992 and references therein) or violent instabilities developing
in presence of small vertical component of magnetic fields
(Balbus \& Hawley, 1992 and references therein)
could play a significant role in angular momentum transport. In future,
we plan to incorporate non-axisymmetry
as well as explicit viscosity to verify some of these assertions.

One of the most significant results of our
simulation is that, for the first time, it is realised that the
winds (a precursor of radio jets ?) with larger kinematic flux
are produced in those cases where strong shocks are also present,
indicating that the shock heating may be an important ingredient in the
ejection of matter from the disk surface. Our simulation also indicates that
a very thorough mixing of matter (not obvious in the figures
we presented here) may be taking place in the post-shock
region before a part of it is expelled from the disk as winds. A
corollary of this process is that, in the event matter undergoes
a significant amount of nucleosynthesis (Chakrabarti, Jin and
Arnett, 1987) in the post-shock region of the disk
(particularly valid in low mass X-ray binaries), the wind
may be rich in metallicity and contribute significantly to the
metallicity of the galaxy which is observed. This is clearly an
important problem, and we plan to pursue this in the near future.

\newpage

\centerline {\large REFERENCES}

\noindent Balbus, S.A. \& Hawley, J.F. 1992, ApJ, 400, 610\\
Chakrabarti, S.K. 1989. ApJ, 347, 365\\
Chakrabarti, S.K., {\it Theory of Transonic Astrophysical Flows}, 1990a,
World Scientific Publ. Co. (Singapore)\\
Chakrabarti, S.K., 1990b, MNRAS, 243, 610\\
Chakrabarti, S.K., 1990c, ApJ, 362, 406\\
Chakrabarti, S.K., Jin, L. \& Arnett, W.D., 1987, ApJ, 313, 674\\
Chakrabarti, S.K. and Molteni, D., ApJ, 1993 (Nov. 10th Issue)\\
Eggum, G.E., Coroniti, F.V. \& Katz, J.I., 1987, ApJ, 323, 634\\
Eggum, G.E., Coroniti, F.V. \& Katz, J.I., 1988, ApJ, 330, 142\\
Gingold, R.A. \& Monaghan, J.J., MNRAS, 1977, 181, 375\\
Hanawa, T., 1988, AA, 196, 152\\
Hawley, J.W., Smarr, L. \& Wilson, J. 1984, ApJ, 277, 296\\
Hawley, J.W., Smarr, L. \& Wilson, J. 1985, ApJ Suppl., 55, 211\\
Lucy L.,  Astron. J., 1977, 82, 1013\\
Monaghan J.J., Comp. Phys. Repts., 1985, 3, 71\\
Monaghan J.J., Ann. Rev. Astron. Astrophys., 1992, 30, 543\\
Paczy\'{n}ski, B. \& Wiita, P.J. 1980, A\&A, 88, 23\\
Shakura, N.I., \& Sunyaev, R.A., A\&A, 24, 337\\
Vishniac E. \& Diamond, P.H., 1992, ApJ, 398, 561\\
Wilson, J.R., 1978, ApJ, 173, 431\\

\newpage

{\centerline {\bf FIGURE CAPTIONS}}

\noindent Fig. 1: Parameter space (bounded by three curves and
is lebeled as $\Sigma$) spanned by the specific energy
${\cal E}$ and specific angular momentum ${\lambda}$ of the flow
for which standing shocks may be formed in the analytical 1.5D disk models.

\noindent Fig. 2: Effects of turbulence is seen on the location of the
stable shock for a range of $\alpha$ (as labeled) and specific energy
${\cal E}$. Specific angular momentum $\lambda$ (in units of $2GM/c$)
is chosen to be $1.65$. With the increase in $\alpha$ the shock seems to be
pushed away from the black hole and the range of energy for which
shocks may be formed seems to go up.

\noindent Fig. 3(a-d): Results of a simulation of thick accretion
disks with shocks (seen here at $x\sim 16.6$) in which matter is injected with
angular momentum $\lambda=1.65$ and energy ${\cal E}=0.006$.
Total number of particles are $60,000$. (a) X-Z coordinates of the particles,
(b) Mach number field of the flow close to the shock, (c) contours of
contant Mach number, and (d) contours of constant non-dimensional
temperature. See text for details.

\noindent  Fig. 4(a-d): Result of a simulation of thick accretion disks
with shocks (seen here at $x=14$) for $\lambda=1.625$ and ${\cal E}=0.007$.
(a) Mach number field of the flow, (b) contours of constant Mach number,
(c) contours of constant temperature, and (d) ratio of the
outflow rate and the injection rate of matter as a function of time.
See text for details.

\noindent Fig. 5(a-b) Simulations results of a thick disk with a
high angular momentum ($\lambda=1.80$). (a) Contours of constant
Mach number showing a vertical shock at $x=40$ and (b) Mach number
field showing the formation of oblique shocks which divert flows towards
the black hole.

\end{document}